\documentclass[aps,prb,12pt,showpacs,showkeys]{revtex4-1}

\usepackage{amsmath}   
\usepackage{graphicx}   
\usepackage{verbatim}  
\usepackage{color}    
\usepackage{subfigure}

\begin{document}

\title{Understanding the metamagnetic transition and magnetic behavior of ${\textrm{Ni}}_{48}$${\textrm{Co}}_{6}$${\textrm{Mn}}_{26}$${\textrm{Al}}_{20}$ polycrystalline ribbons\\} 

\author{Rohit Singh}
\email{rohitau88@gmail.com}
\author{Saurabh Kumar Srivastava}
\email{saurabhphyd@gmail.com}
\author{Ratnamala Chatterjee} 
\email{ratnamalac@gmail.com, rmala@physics.iitd.ac.in}
 \homepage{http://web.iitd.ac.in/~rmala/}
\affiliation{Magnetics and Advanced Ceramics Laboratory, Department of Physics, Indian Institute of Technology Delhi, Hauz Khas, New Delhi-110016, India}

\author{Arun K. Nigam}
\email{aknm@tifr.res.in}
\homepage{http://www.tifr.res.in/~dcmpms/index.php/faculty-members.html?id=282}
\affiliation{Tata Institute of Fundamental Research, Homi Bhabha Road, Mumbai-400005, India}

\author{Vladimir V. Khovaylo}
\email{khovaylo@gmail.com}
\affiliation{National University of Science and Technology “MISiS,” Moscow-119049, Russia}

\author{Lajos K. Varga}
\email{varga@szfki.hu}
\affiliation{Research Institute for Solid State Physics and Optics of the Hungarian Academy of Sciences, H-1525 Budapest, P.O.B. 49, Hungary}

\date{\today}

\begin{abstract}

In this work we demonstrate that the polycrystalline ribbons of (${\textrm{Ni}}_{48}$${\textrm{Co}}_{6}$)${\textrm{Mn}}_{26}$${\textrm{Al}}_{20}$ with B2 structure at room temperature show a magnetic behavior with competing magnetic exchange interactions leading to frozen disorders at low temperatures. It is established that by considering the presence of both antiferromagnetic and ferromagnetic sublattices, we can explain the observed magnetic behavior including the metamagnetic transition observed in these samples. From the Arrott plots, the N{\'e}el  temperature of (${\textrm{Ni}}_{48}$${\textrm{Co}}_{6}$)${\textrm{Mn}}_{26}$${\textrm{Al}}_{20}$ is deduced to be ${\sim 170}$ K and the broad ferro to para like magnetic phase transition is observed at ${\sim 200}$ K. Based on N{\'e}el theory, a cluster model is used to explain the presence of ferromagnetic and anti-ferromagnetic clusters in the studied ribbons. Formation of ferromagnetic clusters can be understood in terms of positive exchange interactions among the Mn atoms that are neighboring to Co atoms which are located on the Ni sites.

\end{abstract}

\pacs{81.30.Kf, 64.70.Kd, 75.30.Kz, 75.60.Ej}
\keywords{Heusler alloys, shape memory alloys, metamagnetism, martensitic transformations, Antiferromagnetism, N{\'e}el Temperature, Curie Temperature.}

\maketitle

\section{INTRODUCTION}

Ferromagnetic shape memory (FSM) effect in stoichimetric and off-stoichiometric full Heusler alloys (${\textrm{X}}_{2}$YZ) has been a subject of recent research interest. Some representatives of these magnetically ordered alloys show strong ferromagnetism alongwith a crystallographically reversible, thermoelastic martensitic transformation resulting in the shape memory effect. Magnetic field induced strain (MFIS) (the highest value\cite{sozinov2002crystal} of ${9\%}$ reported), first shown in ${\textrm{Ni}}_{2}$MnGa by Ullakko \textit{et al.}\cite{ullakko1996large} is explained by the rearrangement of martensite variants under the influence of external magnetic fields. If the energy driving the twin boundaries is smaller than the magnetocrystalline anisotropy energy, the rotation of magnetization vector towards the easy axis of the crystal happens along with variant rearrangement and thus results in MFIS. In this mechanism, the driving force is limited by the anisotropy energy. Despite large MFIS and a rapid response, the output stress which can be generated by this mechanism of MFIS is restricted to only a few MPa.\cite{murray20006}

In 2006, Kainuma \textit{et al.}\cite{kainuma2006magnetic} reported magnetic-field-induced shape recovery in a NiCoMnIn alloy through the reverse phase transformation, in which stress levels of approximately 50 times larger (than the above mentioned kind) could be generated. They obtained $3\%$ deformation and almost full recovery of the original shape of the alloy. This deformation behavior was attributed to a metamagnetic phase transition from the antiferromagnetic (or paramagnetic) martensitic (MST)  to the ferromagnetic austenitic (AST) phase. Since then, a large number of reports on these metamagnetic shape memory alloys (MSMA) have been published. Apart from NiCoMnIn, metamagnetic shape memory effect (MSME) has been observed in NiCoMnSn,\cite{kainuma2006metamagnetic,ito2007martensitic} NiMnIn,\cite{oikawa2006effect} NiMnSn,\cite{koyama2006observation} NiCoMnSb,\cite{yu2007magnetic} NiCoMnGa,\cite{fabbrici2009reverse} NiMnGaCu\cite{jiang2009search} and NiCoMnAl\cite{kainuma2008magnetic} alloys. The magnetostructural transformations in this class of alloy systems are known to manifest several interesting thermomagnetic irreversibilities\cite{ito2008kinetic,sharma2007kinetic,xu2010kinetic,acet2011231} and show many interesting functionalities.\cite{krenke2007magnetic,sharma2006large,moya2007cooling}  

Metamagnetism is a term frequently used to describe a large increase in magnetization of a material at a characteristic field depending on temperature. The metamagnetic transitions observed in various materials can depend on the material as well as on the specific experimental conditions. Thus metamagnetic behavior has been associated in literature with antiferromagnets which upon application of a magnetic field undergo a first order phase transition to a state with a relatively large magnetic moment,\cite{stryjewski1977} or a continuous phase transition at a critical point or even crossovers beyond a critical point that do not involve a phase transition at all. In light of this, it is worth noting that in almost all MSMAs a few features that are observed in common are: (i) in the thermomagnetization curves the parent phase shows stronger magnetism than the martensite phase, resulting in a large change in magnetization (${\Delta M}$) during martensitic transformation, (ii) magnetostructural transformations in these alloys are induced by both applied field (\textit{H}) and temperature (\textit{T}), (iii) when the specimen is subjected to cooling in external magnetic field the martensitic transformation temperature is observed to drastically decrease with increasing field and finally (iv) almost all these alloys are known to show kinetic arrest (KA) behavior below a critical temperature. There are several works in the field that address the magnetostructural hysteretic transitions in thermal cycling of these alloys. Also, observation of kinetic arrest\cite{sharma2007kinetic} implying the presence of a metastable glass like magnetic state at low temperatures and observation of exchange bias in many of these alloys have been reported,\cite{jing2009exchange,khan2007exchange,nayak2009observation} again indicating the possibility of the presence of multiple magnetic interactions in the sample. However, to the best of our knowledge no direct study related to the details of antiferromagnetic interactions or to where the N{\'e}el temperature ${T_{N}}$ may be,\cite{acet2011231} have been addressed yet. It is understood that for the exchange bias effect to occur, the position of ${T_{N}}$ in martensitic Heusler should be lower than the ${T_{C}}$ of the alloy, but what is the position of N{\'e}el temperature in these alloys, is a question that still needs to be answered. It is worth pointing out that unlike other NiCoMnZ alloys, the low temperature martensite phase of NiCoMnAl is not mentioned clearly in literature as antiferromagnetic. Instead, Kainuma \textit{et al.}\cite{kainuma2008magnetic} suggest that the addition of Co in NiCoMnAl system not only induces ferromagnetic properties in parent phase but also changes the magnetic properties of the low temperature martensite phase to paramagnetic.  Also, it should be noted from previous studies\cite{srivastava2011systematic,fujita2000magnetic,manosa2003magnetic,acet2002coexisting} that in Ni-Mn-Al systems the magnetic transitions occur from antiferromagnetic (or from a state with mixed ferro- and antiferromagnetic interactions) to the paramagnetic state. Clearly, further investigations are required to understand the magnetic properties of this system. 
	
	In this work we analyse the detailed magnetic characteristics of ${\textrm{Ni}}_{48}$${\textrm{Co}}_{6}$${\textrm{Mn}}_{26}$${\textrm{Al}}_{20}$ alloy ribbons in the temperature range ${2\;{\textrm{K}}\leq T\leq {330\;{\textrm{K}}}}$ and in the field range ${50\;{\textrm{Oe}}\leq H\leq {70\;{\textrm{kOe}}}}$. Using the Arrott plot analysis we set out to estimate the ${T_{N}}$ of this composition.

\section{SAMPLE PREPARATION AND MEASUREMENTS}

\begin{figure}
	\centering
		\includegraphics[width=1.00\textwidth]{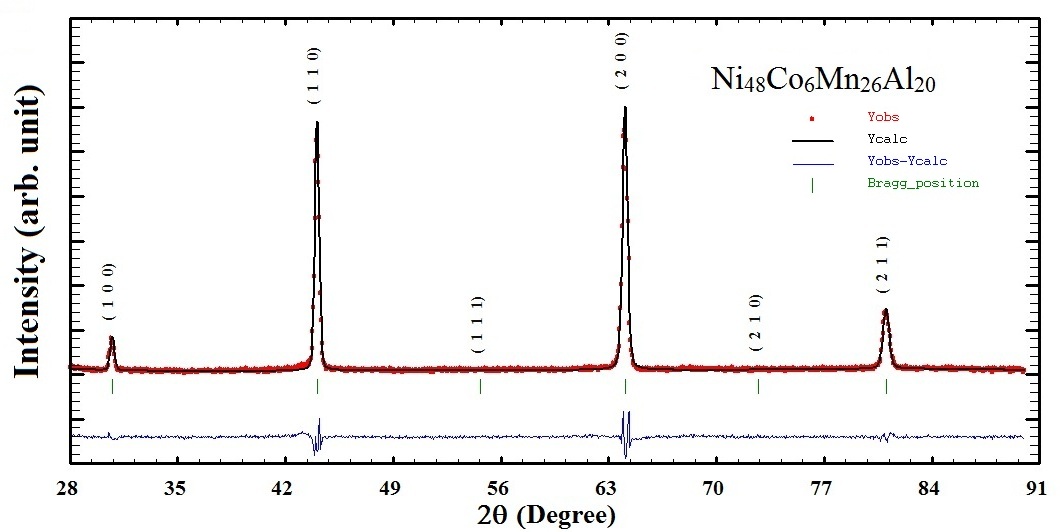}
		\caption{(Color online) XRD pattern of ${\textrm{Ni}}_{48}$${\textrm{Co}}_{6}$${\textrm{Mn}}_{26}$${\textrm{Al}}_{20}$ ribbon samples at room temperature. }
	\label{fig:Figure1}
\end{figure}

	It is known\cite{srivastava2011systematic,ito2008atomic,sutou1998ordering,kainuma2000ordering} that the martensitic transformation temperatures and magnetic properties of stoichiometric and off-stoichiometric ${\textrm{Ni}}_{2}$MnAl-based Heusler alloys strongly depend on the degree of ${\textrm{L2}}_{1}$  surepstructural ordering which is mainly determined by a thermal treatment protocol. In the present case, we have prepared our samples by rapid solidification of the molten alloy in stoichiometric proportion. The ingot was prepared by induction melting in a water-cooled copper boat under argon atmosphere. The mixture of pure Al (${99.99}$ mass ${\%}$), Co (${99.99}$ mass ${\%}$), Mn (${99.9}$ mass ${\%}$) and Ni (${99.99}$ mass ${\%}$) metals were re-melted 5 times to assure homogeneity. Rapidly solidified alloy ribbons with width of ${\sim 1-4}$ mm, thickness of ${20\;\mu}$m  were prepared by melt spinning technique under a partial argon atmosphere. The diameter of the Cu(Zr) roller was 22 cm, with a typical circumferential velocity of 35 m/sec. This process of solidification that is both directional and rapid, leads to a strong crystallographic texture in the ribbon samples, as expected.\cite{dunand2011size} The XRD pattern fitted using LeBail fitting for the Pm3m space group (available with FULLPROF software package) clearly reveals B2 structure at room temperature in these ${\textrm{Ni}}_{48}$${\textrm{Co}}_{6}$${\textrm{Mn}}_{26}$${\textrm{Al}}_{20}$ alloy ribbons (see Fig. 1). The \textit{M} vs \textit{T} measurements were performed using Quantum Design MPMS XL-7 SQUID magnetometer. The samples were first cooled in zero-field down to the lowest temperatures (2 K) and then the required field was applied. These data are denoted as zero-field cooled (ZFC) data. After the sample reached ${\sim 330}$ K, data was recorded during cool down process to lowest temperature of 2 K in field (field cooled cooling or FCC) and then once again the data were collected as the temperature was increased from 2 K to 330 K (field cooled heating or FCH). For \textit{M-H} measurements, sample was first cooled in zero-field down to 5 K and the measurements were made at 5 K for the complete loop. Then the temperature subsequently was raised from 5 K to the desired temperature and the complete loop was measured isothermally at each temperature.

\section{RESULTS AND DISCUSSION}

	\subsection{Magnetization versus Temperature}
	
		\begin{figure}[ht!]
		\centering
		\includegraphics[width=1.00\textwidth,height=12cm]{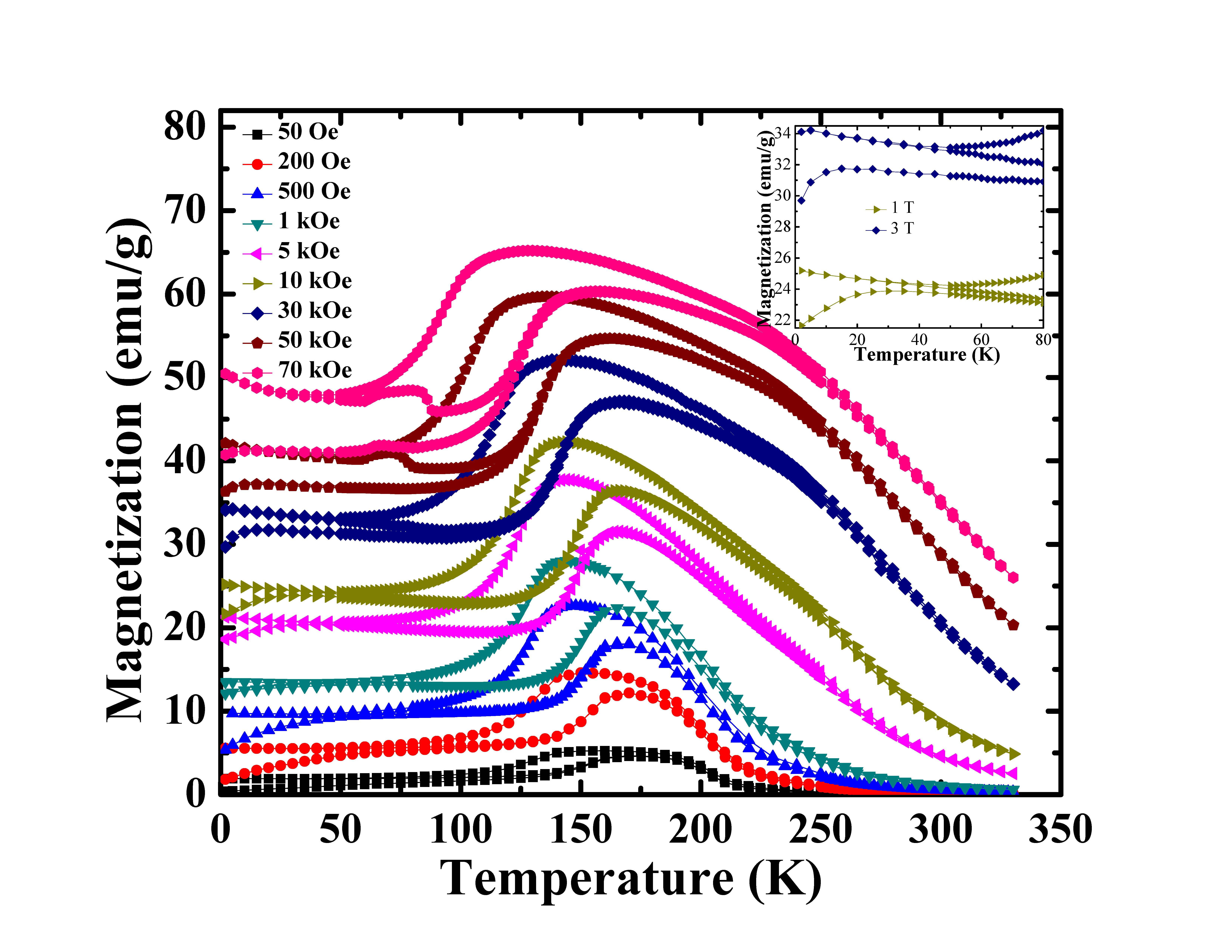}
		\caption{(Color online) Magnetization (\textit{M}) versus temperature (\textit{T}) plots for ${\textrm{Ni}}_{48}$${\textrm{Co}}_{6}$${\textrm{Mn}}_{26}$${\textrm{Al}}_{20}$ sample in external magnetic field \textit{H} = 50, 200, 500, 1 k, 5 k, 10 k, 30 k, 50 k and 70 kOe  over the temperature range (2 K-330 K). Inset shows \textit{M} vs \textit{T} plot for selected fields for temperature range (2 K ${\leq T\leq}$ 80 K).}
		\label{fig:Figure2}
		\end{figure}

	Figure 2 shows the \textit{M} vs \textit{T} (2 K ${\leq T\leq}$ 330 K) curves measured at different fields (50 Oe ${\leq H\leq}$ 70 kOe). Some generic features like thermomagnetic history associated with the magnetostructural phase transitions, shift of martensitic transformation temperatures to lower values on application of magnetic field, large ${\Delta M}$ that increases with increasing field and a large transformation hysteresis of ${\sim 32}$ K as also observed by other researchers\cite{kainuma2008magnetic,xuan2012magnetic} in similar alloy systems, are clearly seen in this sample too. Apart from these, the distinctive features that can be noted in this alloy are (i) a positive slope of ZFC curves at all fields below 50 K and (ii) a signature of a clear field-induced hump noted for ${H\geq}$ 50 kOe between 50 K ${\leq T\leq}$ 75 K. 

	The paramagnetic to ferromagnetic-like transition is observed at ${\sim 200}$ K, which is lower than the values observed for NiCoMnAl polycrystalline bulk samples with similar composition reported by Kainuma \textit{et al.}\cite{kainuma2008magnetic} As the temperature is lowered, a sharp drop in magnetization due to austenite (AST) to martensite (MST) phase transition is noted. A thermal hysteresis between FCC and FCH data is clearly observed around this AST-MST transition, which points out that this transition is of first order. Another interesting feature that is noted in these \textit{M-T} curves at a defined magnetic field \textit{H} is that the magnetization \textit{M} exhibits hysteresis (i.e., the bifurcation between cooling and heating curves) well above the apparent ${T_{C}}$ that was estimated as ${\sim 200}$ K. Evidently, this hysteretic behavior above 200 K is not an experimental artifact (say due to the too fast temperature scan) as the feature appears in high fields but not (or almost not) seen in weak fields. As mentioned above,\cite{srivastava2011systematic} compositional inhomogeneities at nanoscale are not inconceivable in these melt-spun samples, and such compositional inhomogeneities of the sample can result in some martensite to be present even at 250 K and thus can explain the hysteresis of \textit{M} above ${T_{C}}$ as well as the very broad ordered magnetic-to-paramagnetic transition and the large paraprocess of the austenitic phase below ${T_{C}}$.	 
	
	As mentioned above, ${M_{\textrm{ZFC}}}$ shows a positive slope below 50 K (see inset in Fig. 2), whereas, the ${M_{\textrm{FCH}}}$ continues to increase but with a tendency towards saturation below the peak temperature. This irreversible behavior of \textit{M}(\textit{T}) has been regarded in literature as a characteristic feature of spin freezing or spin blocking, wherein the ZFC magnetization is known to show a peak.\cite{bitoh1995field} Morito \textit{et al.}\cite{morito1998magnetic} reported that in Ni-Mn-Al systems, the magnetic state at low temperature is always spin-glass. Also, as pointed out in our earlier work\cite{srivastava2011systematic} on off-stoichiometric Ni-Mn-Al rapidly quenched ribbons, the compositional inhomogeneities at nanoscale result in a spin-glass like phase transition at low temperatures in these melt-spun samples. However, in this case a spin canting, prefiguring of the antiferromagnetism observed at higher temperatures could also be a likely cause.
	
	Apart from this above mentioned broad peak-like feature in ${M_{\textrm{ZFC}}}$ for temperatures below 50 K, it is interesting to note that although the temperature at which the ${M_{\textrm{ZFC}}}$ and  ${M_{\textrm{FCH}}}$ bifurcates (designated here as ${T_{irrev}}$) decreases with increasing field upto 5 kOe, ${T_{irrev}}$ starts to increase with increasing fields above 10 kOe and finally reaches a value of ${\sim 125}$ K for applied field of 70 kOe. The behavior observed in data obtained at fields ${\leq}$ 5 kOe is along expected lines, as in a sufficiently strong external magnetic field the spin freezing either will take place at a lower temperature or will not be seen when the external magnetic fields reach the value which is sufficient to destroy this behavior. However, for the behavior observed at higher fields (for ${H>5}$ kOe) it can be argued that as the applied magnetic field increases, the formation of antiferromagnetic or weakly magnetic martensitic phase gets kinetically arrested and gives rise to a glass-like nonergodic magnetic state. Thus above 30 kOe, the low temperature phase in FCC mode is not an equilibrium phase and contains the converted phase fractions of martensite phase along with the metastable phase fractions of the parent austenite phase resulting in coexistence of ferromagnetic and antiferromagnetic clusters below ${T_{irrev}}$. Indeed, a small hump-like feature in magnetization at ${T\leq{75}}$ K observed for ${H\geq}$ 30 kOe, in the FCC, clearly reinforces the idea of the presence of such multiple phases below ${T_{irrev}}$. Similar behavior has been observed earlier, across the first order ferromagnetic to antiferromagnetic phase transition in doped ${\textrm{CeFe}}_{2}$ alloys\cite{chattopadhyay2005kinetic} and also in ${\textrm{Ni}}_{50}$${\textrm{Mn}}_{34}$${\textrm{In}}_{16}$.\cite{sharma2007kinetic}
	
\subsection{Magnetization versus Field}

\begin{figure}[ht!]
\centering
	\begin{subfigure}
			\centering		
  		\includegraphics[width=0.45\textwidth]{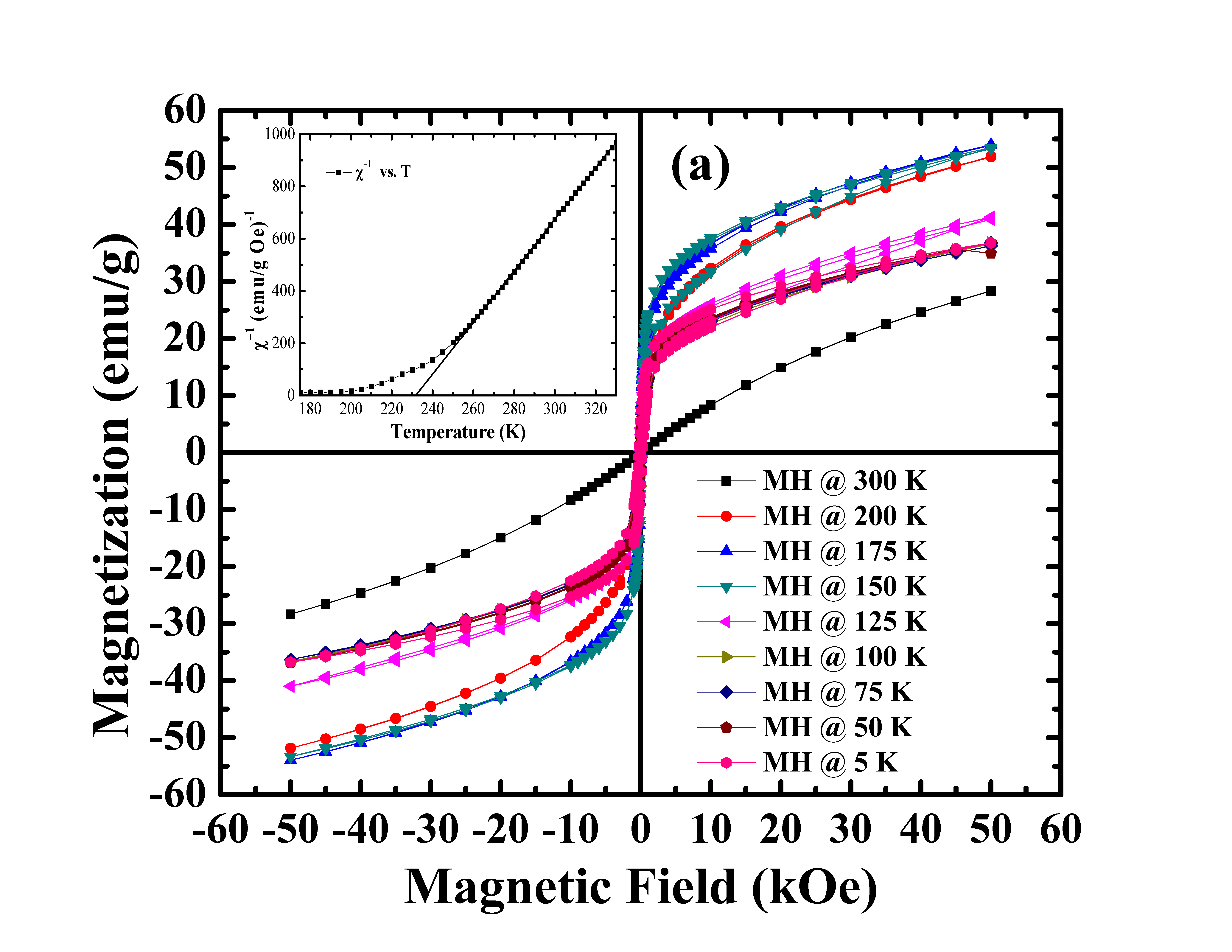}
			\label{fig:3a}
	\end{subfigure}
	\begin{subfigure}
			\centering
			\includegraphics[width=0.45\textwidth]{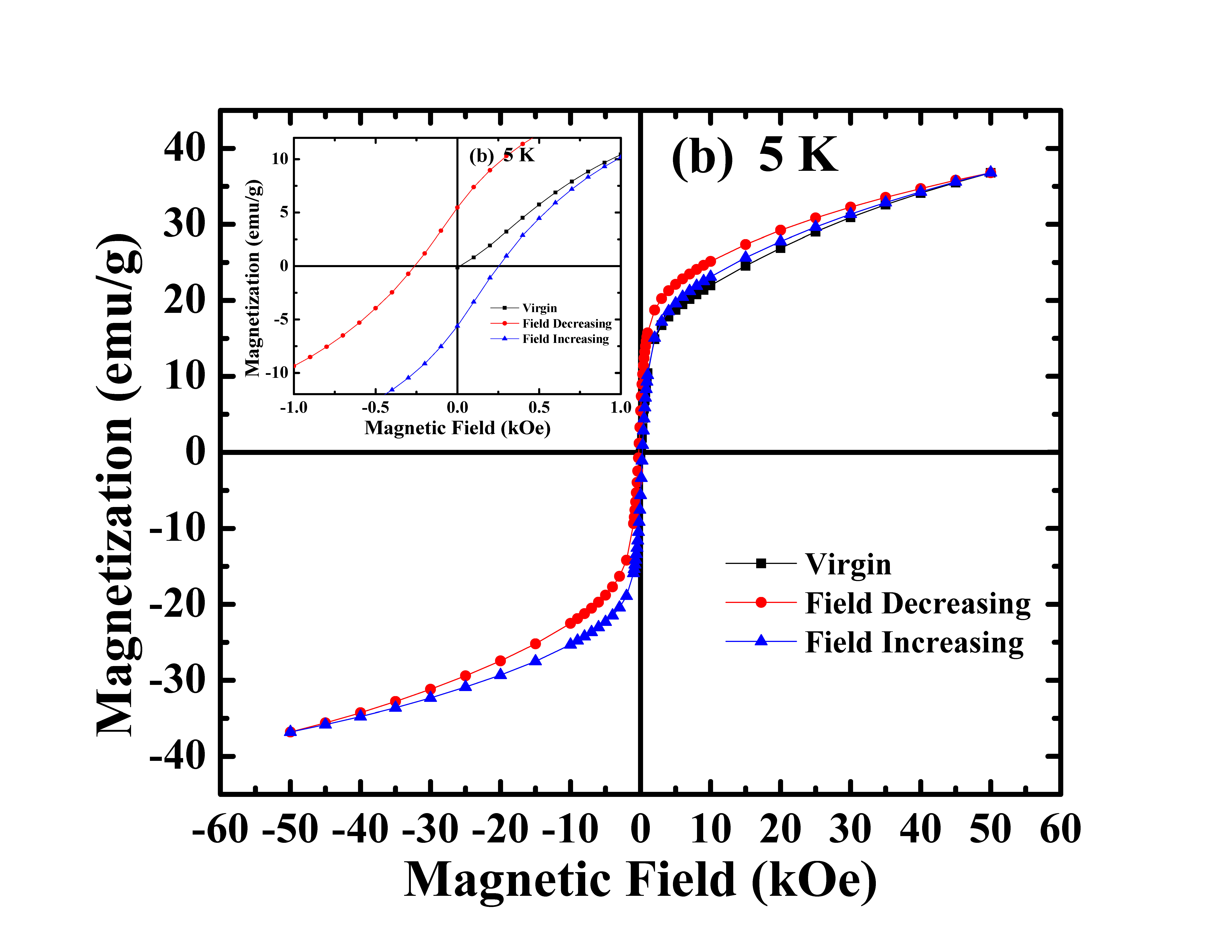}
			\label{fig:3b}
	\end{subfigure}
	\begin{subfigure}
	    \centering
			\includegraphics[width=0.45\textwidth]{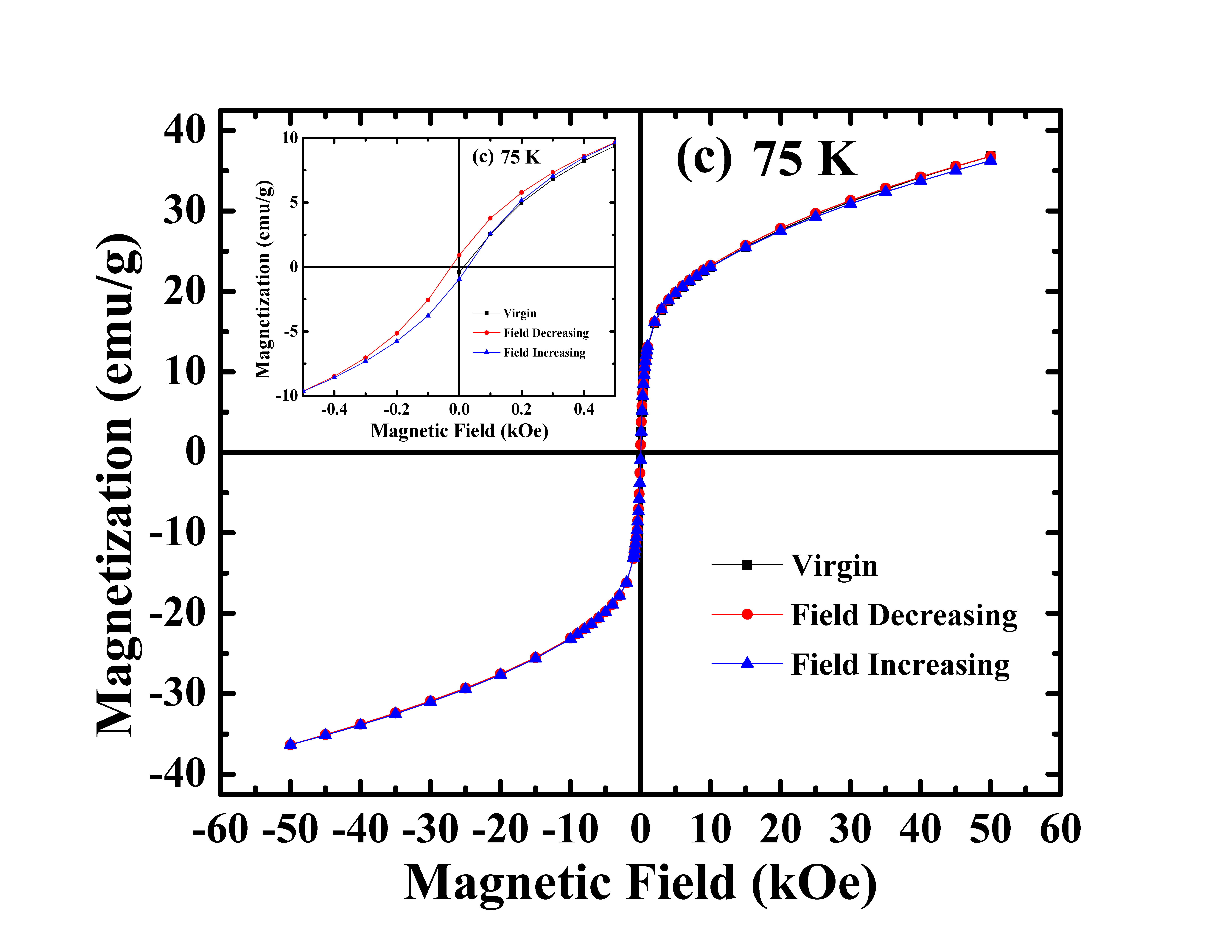}
			\label{fig:3c}
	\end{subfigure}
	\begin{subfigure}
			\centering
			\includegraphics[width=0.45\textwidth]{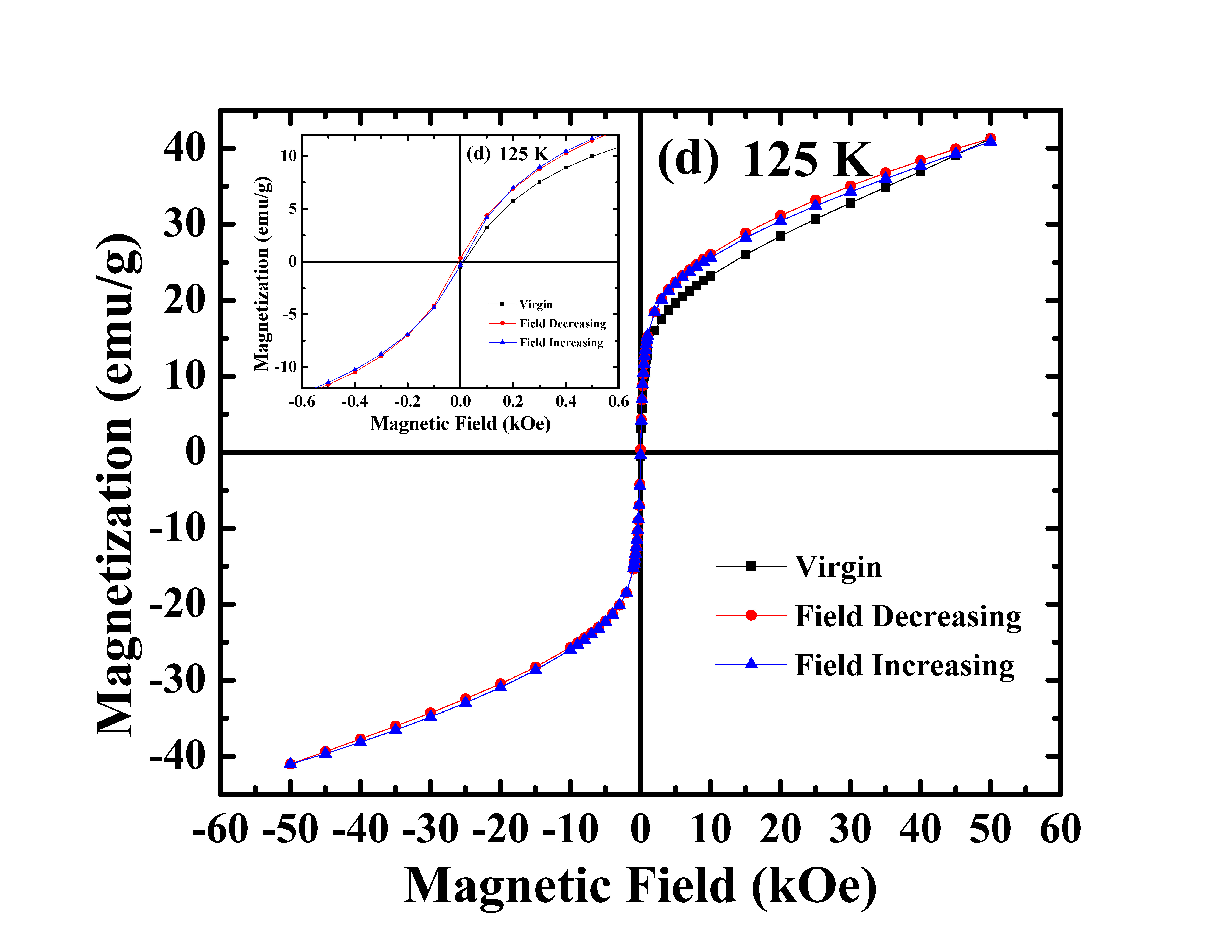}
			\label{fig:3d}
	\end{subfigure}
	\begin{subfigure}
			\centering
			\includegraphics[width=0.45\textwidth]{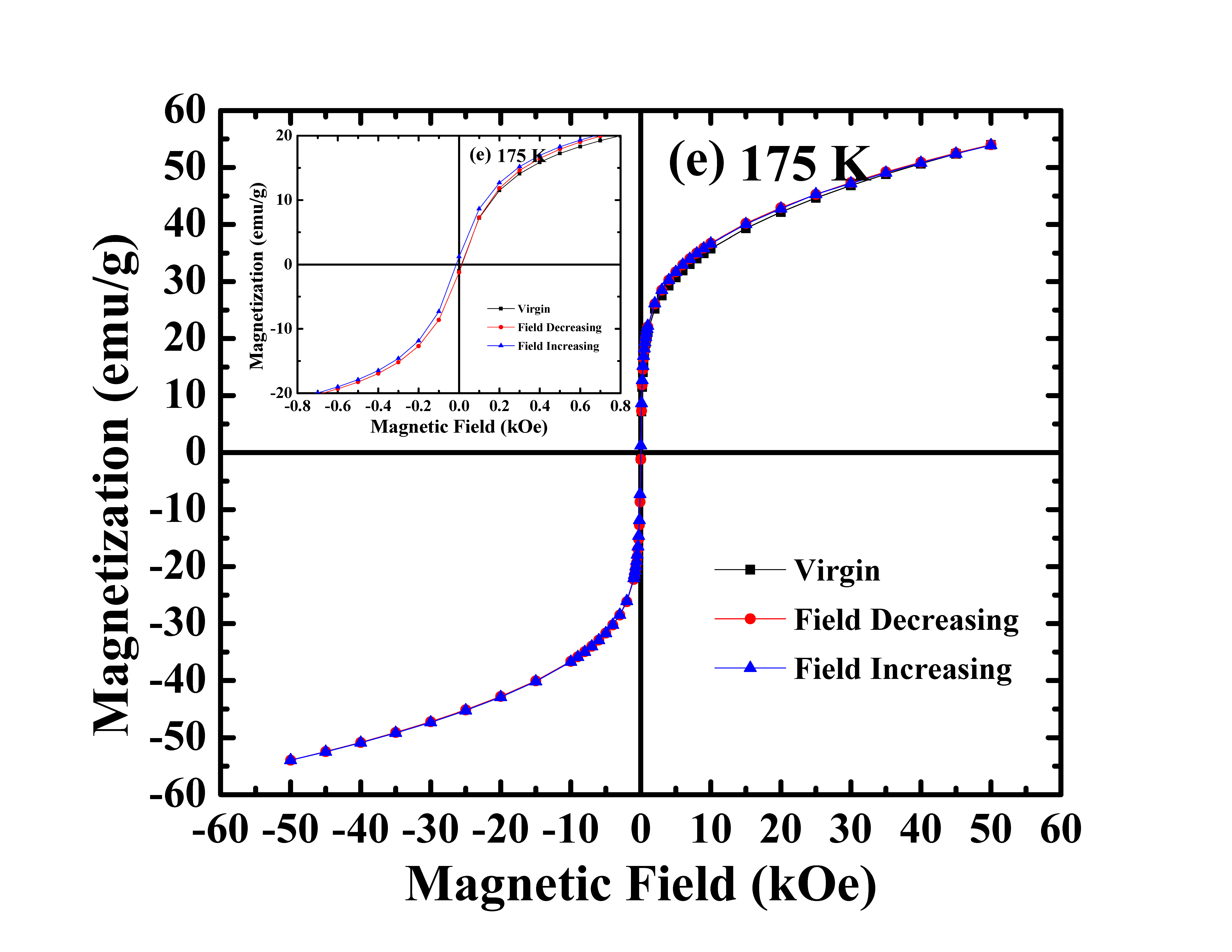}
			\label{fig:3e}
	\end{subfigure}
	\caption{(Color online) (a): Isothermal magnetization (\textit{M}) versus field (\textit{H}) plots for ${\textrm{Ni}}_{48}$${\textrm{Co}}_{6}$${\textrm{Mn}}_{26}$${\textrm{Al}}_{20}$ at various temperatures from 5 K to 300 K. (b-e): Isothermal magnetization (\textit{M}) versus field (\textit{H}) at 5 K, 75 K, 125 K and 175 K respectively. Inset (a): ${\chi^{-1}}$ vs \textit{T} curve for ${\textrm{Ni}}_{48}$${\textrm{Co}}_{6}$${\textrm{Mn}}_{26}$${\textrm{Al}}_{20}$. Inset (b-e): \textit{M} vs \textit{H} curve zoomed near origin to show the coercive field and virgin curve with respect to envelope curve.}
	\label{fig:3}
\end{figure}

The magnetization curves as a function of field (- 50 kOe ${\leq H \leq}$ + 50 kOe) in a temperature range from 5 K to 300 K are shown in Fig. 3(a). The \textit{M} vs \textit{H} dependence taken at 300 K is almost linear close to a paramagnetic state. This agrees well with the results of \textit{M} vs \textit{T} measurements (Fig. 2) which showed that Curie temperature ${T_{C}}$ of the sample studied is well below room temperature. For the field dependencies of the magnetization measured at 5 K ${\leq T\leq}$ 200 K the steep increase of \textit{M} in low magnetic fields is due to a significant role of ferromagnetic interactions in the system. Similar behavior in \textit{M-H} loops has been observed in ferrimagnetic systems also. To ascertain the nature of magnetic interaction, inverse susceptibility has been plotted [see inset in Fig. 3(a)]. The inverse susceptibility for a ferrimagnetic system exhibits a hyperbola shape which becomes more pronounced as the temperature decreases to the magnetic transition temperature.\cite{kaneyoshi2011clear} However, in the present case inverse susceptibility exhibits a linear increase with temperature, confirming the presence of ferromagnetic interaction. In the temperature interval 5 K ${\leq T<}$ 125 K magnetization saturation weakly depends on the temperature and the \textit{M} vs \textit{H} largely overlap each other. It is seen from \textit{M} vs \textit{H} dependence taken at 5 K [Fig. 3(b)] that the virgin curve is lying well within the envelope of the loop; but the slope of the virgin curve is clearly smaller than the slope of \textit{M} in the return loop. The system can be considered as a soft ferromagnet, judging by a rather small coercive field of ${\sim 250}$ Oe [see inset in Fig. 3(b)]. The coercive force becomes even smaller at higher temperatures [see insets in Fig. 3(c) - 3(e)]. The system demonstrates an uncommon feature in the magnetization process at \textit{T} = 125 K [Fig. 3(d)]. Specifically, the magnetization curve measured upon the first application of the magnetic field is located below the magnetization curves taken upon subsequent demagnetization and magnetization of the sample. This feature is easy to understand considering that at \textit{T} = 125 K the sample is in the phase transformation region (Fig. 2) where martensitic and austenitic phases coexist. Upon the first application of the magnetic field an irreversible magnetic field-induced conversion of a part of the martensite fraction with a lower magnetization into austenite with a higher magnetization takes place and brings about the observed feature of the magnetization process.

\subsection{Magnetization and Arrott Plots}

The features observed in Figs. (2) and (3) could arise from either: (i) ferrimagnetic interactions that has been evidenced in many Heusler alloys, or could be resulting from the presence of (ii) multiple magnetic interactions in the sample, as argued in sections A \& B above. It is also clear that in the sample studied, we cannot consider the martensite as purely antiferromagnetic or the austenite as pure ferromagnetic phase. We thus need to address the question whether there are AFM interactions in the austenite phase which get enhanced in the martensitic phase or the austenitic phase is purely (or almost purely) FM.
	
The use of Arrott plots to analyse the bulk magnetic properties of materials is well known in literature. Equation for the Arrott plot,\cite{pjwebster1988heusler,kittel1996introduction}		
	
	\begin{equation}
	M^{2} = \frac{1}{C}\left\{\frac{(B_{0})_{ext}}{M}\right\}-\frac{\tilde{A}}{C}
	\end{equation}
	
resembles the equation for a straight line; plotting the square of magnetization on the y-axis and the ratio of applied magnetic field divided by the observed magnetization on x-axis will yield straight lines. According to the above equation, the lines have a slope proportional to \textit{1/C} and they intersect y-axis at $(-\tilde{A}/C)$, the coefficient $\tilde{A}$ being related to the inverse susceptibility.

	However, since we expect multiple magnetic interactions in our sample, we consider the presence of both antiferromagnetic moment (\textit{L}) and ferromagnetic moment (\textit{M}) in the magnetic sublattice. The expression of the free energy\cite{pjwebster1988heusler} for magnetic sublattice then may be written in the form,
	\begin{equation}
	F=\frac{1}{2}\tilde{A}{M^{2}}+\frac{1}{4}C{M^{4}}- {B_{0}}M + \frac{1}{2}\tilde{a}{L^{2}}+\frac{1}{4}cL^4+\frac{1}{2}\alpha {L^{2}}{M^{2}} + \beta{({LM})^{2}}
	\end{equation}
The terms with proportionality constant $\alpha$ and $\beta$ are the lowest order coupling terms between \textit{M} and \textit{L} allowed by symmetry. The term proportional to $\beta$ determines the orientation of the AFM moment with respect to the ferromagnetic one, and may be incorporated into the term $\alpha {L^2}{M^2}$ with a renormalized coefficient $\gamma$ and then minimizing the free energy in eq. (2) with respect to both \textit{M} and \textit{L} one obtains

	\begin{subequations}
		\label{allequations}
			\begin{eqnarray}
\tilde{A}M+CM^3-B_0+\gamma L^2 M &=& 0, \label{equationa}
\\ 
\tilde{a}L+cL^3+\gamma LM^2 &=& 0, \label{equationb}
			\end{eqnarray}
	\end{subequations}

where, ${B_0} = H$ (applied external magnetic field)
\\
The coefficients $\tilde{A}$ and $\tilde{a}$ are assumed to depend on temperature as

	\begin{subequations}
		\begin{center}
		${\tilde{A} = A'(T-{T_{C}})}$,    ${\qquad\tilde{a} = a'(T-{T_{N}})}$
		\end{center}
	\end{subequations}
	
whereas $C$, $c$, $A'$ and $a'$ are considered to be temperature independent and $A'$ and $a'$ are positive. If the temperature of the system is lowered and only one subsystem is considered, ${T_{N}}$$({T_{C}})$ is the temperature for which the subsystem becomes unstable towards antiferromagnetic (ferromagnetic) order. The instability occurs at the temperature for which the $\tilde{A}$ or $\tilde{a}$ coefficients change sign as a function of temperature. The renormalization of both ferromagnetic ($\tilde{A}$ coefficient) and antiferomagnetic ($\tilde{a}$ coefficient) interactions are seen in the magnetic susceptibility as measured in magnetization experiments. The properties of Arrott plots (${M^{2}}$ vs \textit{H/M}) for a system with ferromagnetic and antiferromagnetic phase transitions are also deduced accordingly.\cite{pjwebster1988heusler}

\begin{figure}[ht!]
\centering
\begin{subfigure}
  \centering
  \includegraphics[width=.45\textwidth]{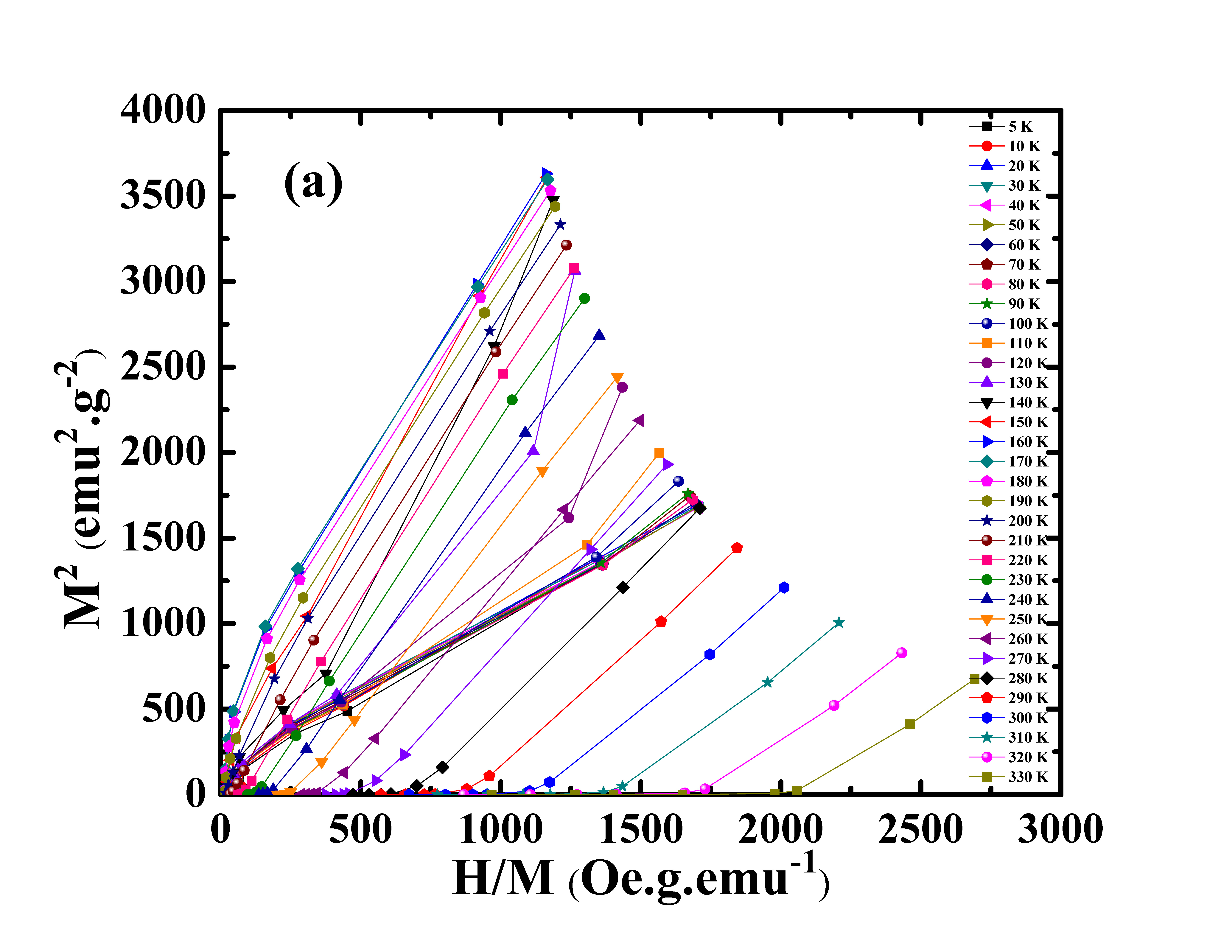}
  \label{fig:4a}
\end{subfigure}
\begin{subfigure}
  \centering
  \includegraphics[width=.45\textwidth]{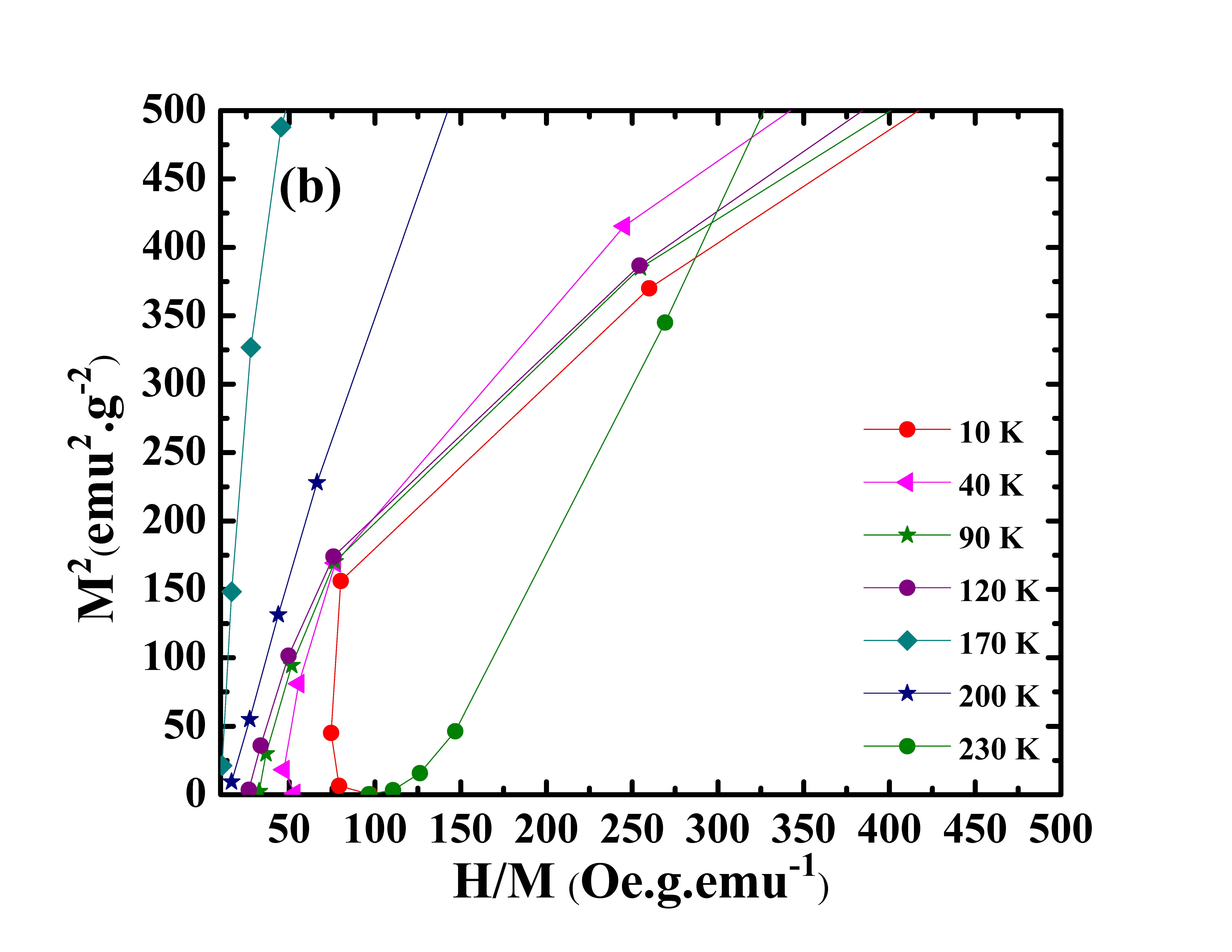}
  \label{fig:4b}
\end{subfigure}
\caption{(Color online) (a) Arrott plots for ${\textrm{Ni}}_{48}$${\textrm{Co}}_{6}$${\textrm{Mn}}_{26}$${\textrm{Al}}_{20}$ at different temperatures ranging from 5 K-330 K. (b) Arrott plots zoomed near origin for selected temperatures, for clearer view.}
\label{fig:4}
\end{figure}
	
	The Arrott plots for our sample at different temperatures (5 K ${\leq T\leq}$ 330 K) are shown in Fig. 4(a). Few features that are clearly evident are: (a) the plots are not linear; as is expected even for weak ferromagnets, (b) near origin all curves show some curvature- for ${T\leq}$ 170 K a negative curvature and for ${T >}$ 200 K a positive curvature was noted, (c) the high field data appear linear, (d) although for \textit{T} = 200 K the line almost seems straight and passing through zero, there is no evidence of spontaneous magnetization at any temperature and most intriguingly, (e) with increasing temperatures, 5 K ${\leq T\leq}$ 170 K the plots are displaced to the left and then for ${T >}$ 170 K as the temperature is increased, plots start getting displaced to the right.
	 
	All the above mentioned features of Arrott plots point out towards the fact that there is no true ferromagnetic order in this sample. This kind of deviation from linearity of isotherms towards the origin in Arrott plots have been observed in spin-glass systems like ${\textrm{Gd}}_{37}$${\textrm{Al}}_{63}$,\cite{malozemoff1983critical} ${\textrm{Y}}_{1-x}$${\textrm{Fe}}_{x}$\cite{coey2000amorphous} or concentrated spin glass system of ${\textrm{Al}}_{37}$${\textrm{Mn}}_{30}$${\textrm{Si}}_{33}$ quasicrystals\cite{chatterjee1990concentrated} near the ferromagnet spin-glass critical concentration. Aharony \& Pytte\cite{aharony1980infinite} predicted in their theory that the deviation from linearity of isotherms towards the origin in Arrott plots is an effect of either a random anisotropy or a random field presented in the system. In their theory they assume a system with a positive \textit{J} and a random anisotropy but neglect the exchange fluctuation effects. According to their theory a random anisotropy system will undergo a broad phase transition at temperature ${T_{R}}$ to a very unusual new magnetic state with an infinite susceptibility but no spontaneous magnetization.   
	
	The signature of positive \textit{J} (ferromagnetic) interactions is seen in the magnetization as measured in \textit{M}(\textit{T, H}) curves of Fig. 2. From Fig. 4(a) it is seen that as the temperature increases from 5 K to 170 K, the Arrott plots are displaced to the left and after 170 K the graphs are displacing towards right for increasing temperatures, indicating ${T_{N}}$ to be at ${\sim 170}$ K. The $\gamma$ values deduced from the plots are observed to be positive and vary between 0 ${\leq \gamma\leq}$ 0.24 for  5 K ${\leq T\leq}$ 150 K, indicating a significant interaction between antiferromagnetic and ferromagnetic moments. The decrease in susceptibility below ${T_{N}}$ arises due to renormalization of the $\tilde{A}$ coefficient.\cite{pjwebster1988heusler} For temperatures above ${T_{N}}$ the antiferromagnetic order parameter is zero. With application of external magnetic field a magnetic moment is induced and with \textit{L} = 0, the magnetization as given in eq. (1), gives straight lines. A change of ${T_{N}}$ as a function of applied fields, which does not couple directly to antiferromagnetic order parameter \textit{L}, is actually an evidence for the renormalization of the antiferromagnetic parameter.

\begin{figure}[ht!]
	\centering
		\includegraphics[width=1.00\textwidth]{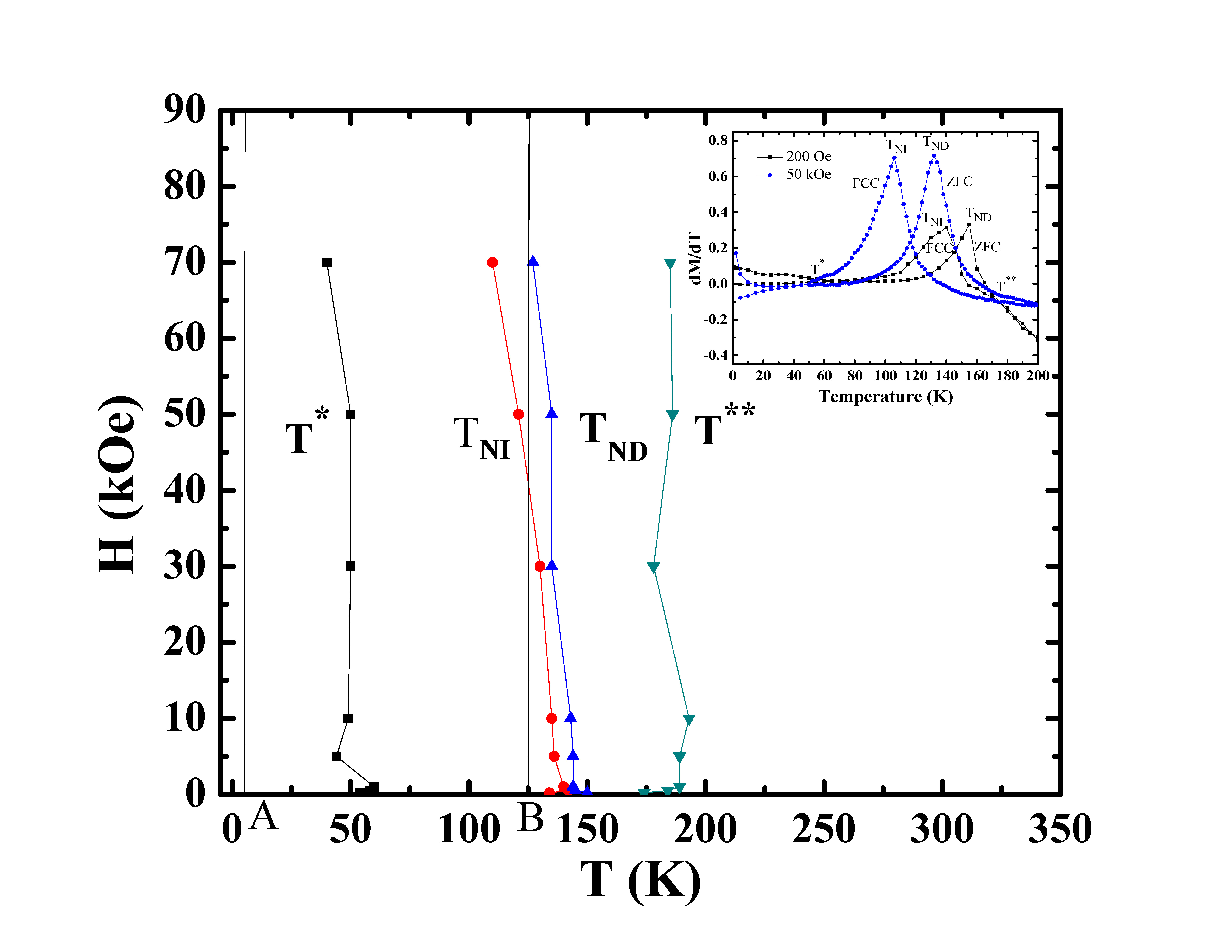}
	\caption{(Color online) \textit{H-T} phase diagram for ${\textrm{Ni}}_{48}$${\textrm{Co}}_{6}$${\textrm{Mn}}_{26}$${\textrm{Al}}_{20}$ samples.}
	\label{fig:5}
\end{figure}
	
	As theoretically argued by Imry and Wortis,\cite{Imry1979influence} such samples with magnetic exchange interactions with frozen disorders at low temperatures show a disorder broadened transition with a spatial distribution of (\textit{H, T}) lines across the sample\cite{chaddah2006studies} on the (\textit{H, T}) phase diagram. We examine a schematic diagram presented in Fig. 5. The lines ${T^{*}}$ and ${T^{**}}$ shown in Fig. 5 are drawn from points at which the hysteresis in \textit{dM/dT} (and in \textit{M} vs \textit{T}) collapses (as estimated from Fig. 1, see inset in Fig. 5). These are the limits of supercooling (${T^{*}}$) and superheating ($T^{**}$) across the transition. The curves ${T_{NI}}$ (or ${T_{ND}}$) are estimated from the sharp rise in the \textit{M-T} curves while increasing (or decreasing) the temperature at a constant \textit{H}. Thus the $T^{**}$ line is depiction of a boundary such that at any temperature above this line, the major portion of the sample is expected to be in the equilibrium austenite FM-like state and for ${T < T^{*}}$ the sample mainly consists of the MST phase along with some metastable phase fractions of the AST phase. Co-existance of the AFM and FM phases could be observed in a temperature intervals ${T^{*} < T < T_{ND}}$ while cooling, and ${T_{NI}< T < T^{**}}$ while heating. The large window of $T^{**}$-$T^{*}$ is the measure of wide transition hysteresis in \textit{M-T} curves observed for this sample. 
	
	Estimated ${T_{N}}$${\sim 170}$ K from Arrott plots along with this phase diagram clearly indicates that the antiferromagnetic behavior persists till very close to the apparent ${T_{C}}$${\sim 200}$ K of the sample. Thus giving rise to the notion that such interactions are possibly present well in to the austenitic phase and then get enhanced in the martensitic phase.
	
\subsection{Formation and coupling of magnetic moments}

	Recent results by Xuan \textit{et al.}\cite{xuan2012magnetic} and Feng \textit{et al.}\cite{feng2012magnetic} on Ni(Co)MnAl alloys point out that Co is the ferromagnetic activator in this alloy system and leads to the ferromagnetic alignment of Mn moments in the AST phase. The detailed energy band calculations in Heusler alloys by K{\"u}bler \textit{et al.}\cite{kubler1983formation} had shown that the states lying well below Fermi level ${E_{F}}$ decide the energy balance in these alloys that determine the magnetic order of the sample. Their work emphasises the importance of \textit{p-d} hybrid states in the vicinity of ${E_{F}}$ and point out the role that these states play in controlling the moment loss associated with alignments other than ferromagnetic. In view of this, the effect of lattice constant (from the variation of Co and Ni content) and the number of non-\textit{d} electrons contributed by Al content are the determining factors for the moment alignment in these alloys. Our XRD results at room temperature lead us to conclude that the structure of the high-temperature AST phase of this sample is cubic B2, in agreement with the previously reported structure for similar alloys by Kainuma \textit{et al.}\cite{kainuma2008magnetic} Also, it is generally accepted that the antiferromagnetism of Ni-Mn-Al is associated with the disordered B2 structure, whereas for ferromagnetic ordering of the magnetic moments the ${\textrm{L2}}_{1}$ ordering of the parent phase is required. In their ab-initio calculation of structure and lattice dynamics in Ni-Mn-Al alloys, B{\"u}sgen \textit{et al.}\cite{busgen2004ab} summarize that the magnetic moment is located mostly at the Mn atomic sites but Ni atoms are responsible for the positive \textit{J} coupling of Mn moments in the Heusler structure. Moreover, the results of Xuan \textit{et al.}\cite{xuan2012magnetic} suggest that the Mn atoms that are neighboring Co atoms, get ferromagnetically coupled with each other through the intermediation of Co. Thus it seems that in this B2 type alloy sample, the antiferromagnetically aligned Mn moments get tuned to the ferromagnetic-like ordering locally near Co neighbor. Thus, if we can think of our sample in the framework of N{\'e}el theory of fine particles\cite{neel1949theorie} then magnetization may be explained in terms of two competing phenomena. We can thus model our system as clusters of spins strongly coupled within the cluster and the clusters as units are coupled to a weak random field. At very high temperatures only paramagnetism exists.  As the temperature is lowered, at ${\sim 220}$ K, the spins within each cluster couple with each other and the moment continues to build, while the various clusters still remain decoupled. At ${\sim 200}$ K, the moment within each cluster increases like a Brillouin function and these cluster moments show a strong response to the external magnetic field. However as the temperature is further reduced, as seen in Fig. 4(a), the antiferromagnetic interactions are favored, and below 170 K the competing interactions lead to frustration in the system and result in a very low net magnetization. On further cooling, the martensitic phase formation gets kinetically arrested; the low temperature phase contains the converted phase fractions of MST along with the metastable phase fractions of the AST phase resulting in coexistence of ferromagnetic and antiferromagnetic clusters and a glass like nonergodic magnetic state below ${T_{irrev}}$.
 
\section{CONCLUSION}

We have established in this work that the polycrystalline ribbons of (${\textrm{Ni}}_{48}$${\textrm{Co}}_{6}$)${\textrm{Mn}}_{26}$${\textrm{Al}}_{20}$ with B2 structure at room temperature show a magnetic behavior with competing magnetic exchange interactions leading to frozen disorders at low temperatures. The features observed in temperature and field dependencies of magnetization can be explained in the realm of our proposed model in which presence of antiferromagnetic and ferromagnetic sublattices are considered. The N{\'e}el temperature ${T_{N}\sim {170}}$ K, as estimated from Arrott plots, along with the \textit{H-T} phase diagram, clearly validates that the antiferromagnetic behavior persists till very close to ${T_{C}\sim 200}$ K. The AFM interactions are thus present well in to the austenite phase and then get enhanced in the martensite phase. It is also shown that in such a sample with two competing magnetic sub-lattices, the magnetization below ${T_{N}}$ decreases due to renormalization of $\tilde{A}$ coefficient. Thus when the presence of competing positive and negative exchange interactions in these polycrystalline ribbon samples are assumed, we can explain the metamagnetic behavior (large high field susceptibility of the austenitic phase, hysteresis of magnetization appearing in strong magnetic fields in the vicinity of ${T_{C}}$) observed in these polycrystalline Huesler alloy ribbon samples. 

\section{Acknowledgements}

This work was supported by DST-RFBR Project (INT/RFBR/P-119 and \#12-02-92700). The DST National SQUID facility at IIT(D), India, is also acknowledged. R.S. would like to acknowledge CSIR, India, for providing a fellowship. R.C would like to thank Prof. Ralph Skomski for his critical reading of the manuscript and help in details.

\bibliographystyle{apsrev}
\bibliography{References}

\end{document}